\documentclass[useAMS,usenatbib,usegraphicx]{mn2e}

\usepackage[dvips]{graphicx}
\usepackage{times}

\let\sec=\section
\let\ssec=\subsection
\let\sssec=\subsubsection
\def\erf{{\rm erf}\,}
\def\japref{\parskip =0pt\par\noindent\hangindent\parindent
 \parskip =2ex plus .5ex minus .1ex}
\def\[{\begin{equation}}
\def\]{\end{equation}}
\def\prob{P}
\renewcommand{\theenumi}{(\arabic{enumi})}

\title[Bayesian evidence in astronomy]{The power of Bayesian evidence in astronomy}

\author[Jenkins \& Peacock]{C. R. Jenkins$^{1}$\thanks{charles.jenkins@csiro.au}
and J. A. Peacock$^{2}$\\
$^{1}$CSIRO Earth Sciences and Resource Engineering, Pye Laboratory, Clunies Ross Drive, Canberra ACT 2601 Australia\\
$^{2}$Institute for Astronomy, University of Edinburgh, Royal
Observatory, Blackford Hill, Edinburgh EH9 3HJ}

\begin{document}

\date{Received 2010}

\pagerange{\pageref{firstpage}--\pageref{lastpage}} \pubyear{2010}

\maketitle

\label{firstpage}

\begin{abstract} We discuss the use of the Bayesian evidence ratio, or
Bayes factor, for model selection in astronomy. We treat the evidence
ratio as a statistic and investigate its distribution over an ensemble
of experiments, considering both simple analytical examples and some
more realistic cases, which require numerical simulation.  We find
that the evidence ratio is a noisy statistic, and thus it may not be
sensible to decide to accept or reject a model based solely on whether
the evidence ratio reaches some threshold value.  The odds suggested
by the evidence ratio bear no obvious relationship to the power or
Type I error rate of a test based on the evidence ratio.  The general
performance of such tests is strongly affected by the signal to noise
ratio in the data, the assumed priors, and the threshold in the
evidence ratio that is taken as `decisive'. The comprehensiveness of
the model suite under consideration is also very important. The
usefulness of the evidence ratio approach in a given problem can be
assessed in advance of the experiment, using simple models and
numerical approximations. In many cases, this approach can be as
informative as a much more costly full-scale Bayesian analysis of a
complex problem.
\end{abstract}


\sec{INTRODUCTION}

The apparatus of Bayesian evidence has  been proposed as the
preferred means of answering questions concerning model complexity
in astronomy (e.g. Trotta 2008). Astronomers commonly wish to decide whether a
given model fits a dataset adequately, or whether there is a need
for additional degrees of freedom. Bayesian methods are attractive
in this context because they expose any assumptions or prior
information being used, and permit a clear statement of the
questions that scientists actually ask of their data (e.g. Jaynes
2003). They have been used extensively in the difficult
problems of inference that arise in cosmology (e.g. Hobson et al. 2010), and also in the
complex, multi-parameter modelling needed for the discovery of
exoplanets (e.g. Gregory 2005). Astronomy is by no means
the only area of  science where these methodological questions are
posed or where Bayesian methods are proposed as the solution;
but the astronomical literature on the topic
raises some  issues  worth treating in context.

Bayesian methods give a transparent framework for model choice, in
which it is necessary
to define the set of competing models explicitly and exhaustively;
Bayes' theorem then gives the probability of any particular model
being correct. Integrating this Bayesian probability over the
parameter space associated with the models (as detailed below in
Section 2) then yields an overall ratio of odds for particular
classes of models: the `evidence ratio'. The requisite multi-dimensional integrations
over the parameter spaces of the models
present a  computational challenge, and astronomers have
made  contributions to the development of these
techniques. The refinement of Markov Chain Monte Carlo methods for
the evaluation of multi-dimensional integrals is an example
(Skilling 2004; Mukherjee, Parkinson \& Liddle 2006; Feroz, Hobson
\& Bridges 2009). Numerical Bayesian are establishing themselves
as a default approach, via public-domain packages such as
CosmoMC ({\tt http://cosmologist.info/cosmomc/}) and
MultiNest ({\tt http://ccpforge.cse.rl.ac.uk/gf/project/multinest/}).

Accepting the evidence ratio methodology, some authors have gone
further and attempted to place a quality measure on future experiments
according to the expected evidence ratio values that they are
predicted to yield for given decision problems (Trotta 2007b; Heavens,
Kitching \& Verde 2007). What has been missing from this discussion,
however, is an assessment of the statistical {\it power\/} of the
evidence ratio: different realizations of data for a given
experimental configuration will yield different values of the evidence ratio, and we need to know how often the method will discriminate
correctly between models, and how often it will fail.  This is a
frequentist view of a Bayesian tool, but there is no conflict: the
evidence ratio is a statistic generated from a dataset, so
it is legitimate to ask how it will behave under repeated trials.

We will discuss several examples where the evidence ratio can be
used for model choice, and we will examine the statistical variation
that results from different realizations of the data. The variation
is considerable, and we  argue that this is likely to be generally
true. This suggest caution in the use of evidence
ratios, but it suggests that simplified methods can be
used to compute evidence ratios and check their robustness.

Notwithstanding these caveats, we do advocate a more widespread use of
the evidence ratio technique in astronomy. Bayesian methods are
currently usually employed on complex, high-value
problems; but astronomers are also interested in simpler model choice
problems where the Bayesian techniques have much to offer and are much
easier to use (at least in an approximate way). It is feasible to
experiment with these simpler cases and get a good sense of the
robustness of the method. Approximate Bayesian methods may often be as
good as is justified by the data.

\sec{THE BAYESIAN EVIDENCE RATIO METHOD}

Suppose we have just two models $H_0$ and $H_1$, associated with
sets of parameters $\vec{\alpha}$ and $\vec{\beta}$. For data $D$,
Bayes' theorem  gives the posterior probabilities of the models and
their parameters:
\[
\prob(H_0,\vec {\alpha} \mid D) \propto \prob(D \mid H_0,\vec {\alpha}) \times
\prob(\vec{ \alpha} \mid H_0) \times \prob(H_0)
\]
and
\[
\prob(H_1,\vec{\beta} \mid D) \propto \prob(D \mid H_1,\vec{\beta})\times
\prob(\vec{\beta} \mid H_1) \times \prob(H_1).
\]
Here the priors are, for instance, $\prob(\vec{\alpha} \mid H_0)$,
the probability distribution of the parameters given model $H_0$,
multiplied by the prior probability of the model {\it class\/} $H_0$ itself. We
can often avoid  the (common) normalizing factor required in these
equations. It divides out whenever we take the ratio to form
relative probabilities or `odds'.

The restriction to two models is not fundamental.  Often $H_0$ is
the `null' or default hypothesis and is relatively simple and well
understood.  It is vital that $H_1$ be reasonably comprehensive,
covering a range of possibilities, as otherwise the evidence ratio
formalism may result in high odds in favour of one of the models
when both are a poor fit.

The term `model' can commonly be applied to each distinct point in
parameter space, but a distinct question is how reasonable a given
{\it class\/} of model is in the face of some data. When we discuss
`model selection', we are thus interested in the general viability
of $H_0$ or $H_1$, irrespective of the exact value of their parameters. Integrating out the parameters gives the posterior
probabilities of $H_0$ and $H_1$, conditional on the data. The ratio of these
probabilities is the {\it posterior odds}, ${\cal O}$:
\[
{\cal O}
\equiv
\frac{\prob(H_1\mid D)}{\prob(H_0\mid D)}
=
\frac{\int \prob(D \mid H_1,\vec {\beta})\prob(\vec{ \beta} \mid H_1)\,
d\vec{ \beta} }{\int \prob(D \mid H_0,\vec
{\alpha})\prob(\vec{ \alpha} \mid H_0)\, d\vec{ \alpha}
}\times \frac{\prob(H_1)}{\prob(H_0)}.
\]
We assume that our set of possible models is exhaustive, so that
$\prob(H_1\mid D)+\prob(H_0\mid D)=1$, the probability of $H_0$ is
\[
\prob(H_0\mid D)=\frac{1}{1+{\cal O}}.
\]
For more than two models, this does not hold, but ${\cal O}$ always
gives the relative probabilities of any two models.

The odds ratio ${\cal O}$ updates the prior odds on the models,
$\prob(H_1)/\prob(H_0)$, by a factor that depends on the data:
\[
({\rm Posterior\ odds}) = ({\rm evidence\ ratio}) \times ({\rm prior\ odds}),
\]
or
\[
{\cal O} = {\cal E} \times {\cal PO},
\]
where the definition of the evidence ratio ${\cal E}$ involves
integrals over the likelihood function times the priors on
the parameters:
\[
{\cal E} = \frac{\int {\cal L}(\vec {\beta}\mid H_1)\,
  \prob(\vec{ \beta} \mid H_1)\, d\vec{ \beta} }
{\int {\cal L}(\vec {\alpha}\mid H_0)\, \prob(\vec{ \alpha} \mid H_0)
  \, d\vec{ \alpha} }.
\]
The priors have to be properly normalized and may be quite different
for $H_0$ and $H_1$. If the models in question are also hard to
calculate, the computational problem is large.

A decision about which model to prefer thus requires both the evidence ratio and the prior ratio. The prior ratio is often taken as unity,
but this is not always  justified. For example, one might be
reluctant to accept (say) $H_1$ with 100 free parameters
if $H_0$ had no free parameters. The evidence ratio contains a different
penalty for unnecessary complexity in the models: models are
penalized if a small part of their prior parameter range matches the
data. This is often called the Ockham `factor' (e.g. p348 of
Mackay 2003), although it is not usually an explicit multiplicative
penalty based on the number of parameters.

In this paper, we will always take the prior ratio to be unity, in the
interests of  brevity.  This allows us in our examples to use
`evidence ratio' and `odds' interchangeably, the latter being often
more illuminating.

The roles of the priors on the parameters, and the Ockham penalty,
have been extensively discussed. Recent examples include Trotta (2008)
and Niarchou, Jaffe \& Pogosian (2004).  In hard problems, the prior
and the likelihood can be of similar importance in determining the
value of the integral, and their product may be multi-peaked or
otherwise pathological.

Many interesting cases are however much easier. The first examples we
will discuss can be solved analytically. More generally, if our data
are informative, the likelihood function may be considerably
narrower than the prior. The priors can then be approximated by
constants over the relevant  range of the parameters in the evidence
integrals. Furthermore, in simple cases the integrand may be close to Gaussian
around its peak, in which case the consequent integration of a
multivariate Gaussian can be done analytically:
\[
\int L(\vec{\alpha})\, P(\vec{\alpha})\, d \vec{\alpha}
 \simeq \frac{(2\pi)^{m/2}}{ \sqrt{\mid\mbox{det}({\cal H})\mid} }\,
L(\vec{\alpha}^*)\, P(\vec{\alpha}^*),
\]
where $\vec{\alpha}^*$ is the value at the peak of the likelihood,
$\cal{ H}$ is the Hessian matrix of second derivatives of the
log of the likelihood at the peak, and $m$ is the number of parameters.
This equation is known  as the Laplace approximation, or the method of
steepest descent (see e.g. p341 of Mackay 2003).

The integration then reduces to the less laborious task of finding the
maximum posterior probability, and evaluating the matrix $\cal{ H}$.
Averaging $\cal{ H}$ over many realizations of the data yields the
Fisher matrix, which may be inverted to yield an approximate prediction for the covariance matrix of the
parameters (e.g. Tegmark, Taylor \& Heavens 1997).

The Laplace approximation may not be valid, since the posterior may
not be Gaussian near its peak, or there may be multiple peaks of
similar height.  The applicability of the approximation thus needs to
be checked, at least via inspection of the posterior, or via
comparison with an alternative robust means of integration, such as
Monte Carlo.  Monte Carlo methods can be also used to quantify the
robustness of the evidence ratio for different realizations of the
data.  In addition to providing possible indications of multimodality
in the posterior, such an approach can also probe the stability of the
evidence ratio against systematic error at plausible levels.

\sec{THE EVIDENCE AS A STATISTIC}

\ssec{Repeated experiments}

Posterior odds on hypotheses can be obtained from a given dataset
without ever having to consider whether the experiment could be
repeated. This is the well-known advantage of Bayesian reasoning over
the frequentist approach. Yet many kinds of experiment can be and are
repeated many times: observations in astronomical surveys are an
obvious example. In these circumstances, the Bayesian evidence is a
statistic that can be computed from a given dataset, and it is hard
not to wonder what value might have been obtained had our dataset been
a different realization of the experimental process. Clearly if we
know the likelihood function, which we must do to compute the
evidence, then we must be able to generate other possible realizations
of the data.

To make a decision based on the posterior odds, a threshold is set
at some value of posterior odds, such as the `decisive' $\ln {\cal
E}=5$ value advocated by Jeffreys (1961). We may ask how often such a strategy might lead us to make an incorrect decision. Conversely, it is useful to know if a given experimental setup is likely to yield data good enough to exceed the
decision threshold and `detect' the more complex model in cases
where it is true.

Suppose further that the evidence ratio turns out to be a `noisy' statistic, in the
sense that its distribution is very broad:
in this case, there is little point in devoting excessive
effort in computing the evidence ratio very precisely. Given that
practical computations can involve difficult integrations
over spaces of very high dimensionality, this is worth knowing.

The notion of `repeated trials' needs clarification. In the simplest case, the fluctuations in our data arise in
the measurement process,  while the object or process we are observing
has fixed parameters.  A distinct case arises when we make repeated
measurements of objects or processes that are different on each
repetition. This often happens when we are observing samples and wish
to make statements about properties of whole populations.  In this
case, extra variance enters, often called cosmic variance.
An elementary example is the distinction between repeated (noisy) measurements of
the flux of a single galaxy, or a series of measurements where a
different galaxy is observed on each occasion. In the latter
case, we will have a prior distribution for the true flux
of a randomly selected galaxy, and the data we obtain in a
given measurement could be modelled by drawing a
random number from this prior distribution, and then adding noise.

In dealing with the evidence ratio for repeated trials, we will thus
use the prior {\it twice\/}.  The standard Bayesian approach regards
the data as being fixed specific numbers, and the prior enters only
when we average the likelihood function over the prior to obtain the
posterior probabilities.  However, when we view the Bayesian outputs
as statistics, we have to treat the data as random variables, whose
distribution will depend on the values of the parameters for which we
have a prior.  The probability distribution of the evidence ratio
involves the data, and so depends on the unknown parameters that are
the argument of the prior.  We can eliminate these parameters by a
further integration over the prior, in effect, marginalizing the
distribution of the evidence ratio to obtain its probability
distribution independent of parameters.

\ssec{Neyman-Pearson Analysis}

Suppose we have the posterior probabilities or posterior
odds for our competing models.  These will vary with
different realizations of the data.  What do we
do with these probabilities or odds? This is not a question that can
be answered by probability theory but it can be illuminated by it.

One approach is to set a threshold in the odds,
effectively taking one decision if our experiment gives posterior
odds above the threshold, and another if they lie below.  This
general idea was introduced to classical statistics by Neyman and
Pearson.  A Bayesian approach is to
emphasize the posterior probabilities or odds as a complete summary
of our state of knowledge after the experiment, and to resist further
interpretation. There are parallels here with the long history of
controversy in classical statistics about the Neyman-Pearson
approach versus Fisher's significance testing.  Fisher recognized the utility of 
the Neyman-Pearson method in industrial acceptance
testing but regarded it as too ``wooden'' to be useful in the
ill-defined and creative processes of science (Fisher 1956).  His detestation of
Bayesian methods aside, Fisher would perhaps have been sympathetic to
the idea that posterior probabilities should be carried forward
intact through the processes of science; he took much the same  view
of the results of his tests of significance.

We  believe that binary choice between alternatives is a
relevant process in astronomy.  The high cost and complexity of many
astronomical research projects requires a difficult decision on when
to commit to construction, which is irrevocable once made. More
generally, there is the whole issue of how a community develops a
consensus. The Bayesian ideal of a set of individuals each
interpreting the evidence ratio in the their own way is hardly
realistic: rather, some pre-defined level of proof is needed -- a
threshold in evidence ratio, in short.  We will therefore apply a
Neyman-Pearson style of analysis to the evidence ratio, despite
recognizing that this is not a unique assessment of is utility.

Given the distribution of the evidence ratio ${\cal E}$ under two
competing hypotheses, we can  ask how well the statistic performs. A
Neyman-Pearson analysis proceeds  by defining a critical threshold in
the test statistic, say $ {\cal E}_c$. If $ {\cal E}<{\cal E}_c$, we
do not see any reason to reject the simpler null hypothesis $H_0$,
and it is accepted. If ${\cal E}>{\cal E}_c$,
there is good reason to prefer the more complex hypothesis $H_1$
and $H_0$ will be rejected. A common `decisive' choice for the
critical threshold is $\ln {\cal E}_c=5$, corresponding to odds in
favour of $H_1$ of 148:1 (Jeffreys 1961; Jaynes 2003). The common restriction to two models is not critical, since we can
always add the posterior probabilities for $N$ alternative models
and consider this to be a single alternative. This yields sensible
answers, even in the case where all $N$ models fit the data about
as well as $H_0$: if $N>148$, we would then decide that there was
decisive evidence against $H_0$, even though $H_0$ fitted as well
as any model. This simply reflects our assumption that all models
are equally likely a priori.

In the Neyman-Pearson approach, there are two ways in which an
incorrect conclusion might be reached:

\begin{enumerate}
\item Type I Error (False Positive). $H_0$ is true, but we are unlucky
enough to get a high value of ${\cal E}$ above ${\cal E}_c$, so $H_0$
is incorrectly rejected.  The Type I error rate is
$\alpha\equiv\prob({\cal E}>{\cal E}_c \mid H_0)$.
\item Type II Error (False Negative). $H_1$ is true, but a value of ${\cal E}$
below the threshold is found, so we fail to `detect' the need for a
more complex model.  The power is 1-Type II error rate and so is
${\cal P}\equiv\prob({\cal E}<{\cal E}_c \mid H_1)$.
\end{enumerate}

\noindent The power of the test is defined as the probability that
we will correctly pick $H_1$ when it is true -- i.e. it is unity minus the
probability of a Type II error. There is the usual trade-off: if we
conservatively use a high threshold, we reduce the chance of a Type
I error, but we also reduce the power of the test because we are
increasing the probability of a Type II error.
The power is less often discussed in astronomy, because
alternative hypotheses are frequently ill-defined.  However, the
notion of these alternatives is inherent in the Bayesian method.

An example might be a case where we calculate a statistic, say
chi-square, to evaluate the goodness of fit of a model $H_0$.  From
this we may compute what is often called the $p$-value, the probability
of exceeding the value of chi-square we actually obtained, assuming
$H_0$ to be correct.  This is classical significance testing.  In the
Neyman-Pearson method, we would fix in advance a critical value of
chi-square, corresponding (say) to $p=0.05$.  If we exceed this
critical level (by any amount) we reject $H_0$; if we only have two
models, we then accept $H_1$.

The Neyman-Pearson binary decision rule avoids the paradoxical issue
of the relation between $p$ and $\prob(H_0\mid D)$, which has been
pondered by a long series of authors (Lindley 1957; Jeffreys 1961 and
others cited in Berger \& Sellke 1987; Sellke, Bayarri \& Berger
2001). Paradoxes arise because we might think that if we obtained a
small value of $p$, it would follow that $H_0$ was unlikely to be
correct. While this is not a rigorous interpretation of the meaning of
$p$, successive authors have calibrated $p$ for a range of models, and
found that $p<\prob(H_0 \mid D)$.  This situation can be understood if
we select outcomes with given $p$ from an {\it ensemble\/} of repeated
experiments in which $H_0$ and $H_1$ are equally likely. At one
extreme, the two models may be rather similar, in which case an
outcome with any $p$ is equally probable on either model; if the
models are extremely different, $H_1$ might always yield $p\ll1$, so
observing e.g. $p=0.05$ could in fact provide strong reason to prefer
$H_0$.  Thus $H_0$ can be more likely than the value of $p$ might seem
to suggest, and a calibration of $p$ is required: without this, it is
hard to know what {\em decisions} to take on the basis of $p$. This
discussion has continued since Fisher introduced $p$ and the idea of
significance testing.

The Neyman-Pearson approach, by contrast, is quite clear about how
statistics should be used to take decisions and it is for this reason
that we analyze the performance of the Bayesian evidence ratio using
the concepts of Type I error rate and power.  Our analytical models
are the same as those discussed before (Lindley 1957; Jeffreys 1961,
for example) and we reproduce the $p<\prob(H_0 \mid D)$ effect.  But this
is not our focus; our interest is in showing how the Bayesian
evidence ratio performs, as a basis for decision, under repeated
trials.

An advantage of the Neyman-Pearson approach is that because there is a clear
decision rule, the risks are also clear.  For example, if there is a
cost of some sort associated with wrongly rejecting the null
hypothesis, then the threshold can be set to minimize this.  As we
shall see, however, this also affects the power -- and may affect the
likely pay-off of correctly choosing the alternative.

The Neyman-Pearson lemma tells us that a statistic based on the
likelihood ratio will be the best one to use as a basis for decision.
The evidence ratio is quite closely related to a likelihood ratio and so
this is another reason for examining its performance from a
Neyman-Pearson point of view.

We also note that the Neyman-Pearson approach can also be assessed in a Bayesian
way: we might ask, what is the probability of (say) $H_1$, given that
I have just obtained ${\cal E}>{\cal E}_c$?  This quantity, $\prob(
H_1 \mid {\cal E}<{\cal E}_c)$, is called the `positive predictive
power' in medical literature.  It is related by Bayes' Theorem to the
Type I and Type II error rates, and the prior odds ratio on $H_1$ and
$H_0$:
\[
{\rm PPP}=\frac{\Theta {\cal P}}{\alpha+ \Theta {\cal P}},
\]
in which $\alpha$ is the false positive rate,
${\cal P}$ is the power, and $\Theta$ is the prior odds on $H_1$.
Evidently the use of the PPP
requires us to set a threshold in advance on our test statistic, much
as in the Neyman-Pearson approach.  We might then pose similar
counterfactual questions such as, if we have obtained ${\cal E}<{\cal E}_c$ and we then choose $H_1$ with probability PPP, what might our
loss be if $H_0$ is true?

\sec{GAUSSIAN EXAMPLES}

We will now examine two contrasting Gaussian examples. In the first, both priors are very
narrow and the evidence ratio acts in the same way as a classical
statistic for model choice.  In the second we have one prior much
wider than the other.  The evidence ratio method now diverges
strongly from a classical alternative and the role of the Ockham
factor is apparent.  In both cases we will see how finite amounts of
data reduce the effectiveness of decisions based on thresholds in the
evidence ratio.

We have $N$ data values, $X_i$, which may have arisen via
one of two models:

\begin{enumerate}
\item Independent drawings from a Gaussian of unit standard deviation and mean zero.
\item Independent drawings from a Gaussian of unit standard deviation and mean $\mu$.
\end{enumerate}

\noindent As usual, we assume that these two possibilities are a priori
equally probable.
This is a case where the two models
are {\it nested\/}: model 1 is a special case of model 2 ($\mu = 0$).

In astronomical terms, this situation might correspond to a
(one-dimensional) Gaussian source in the zero-background limit (X-ray
astronomy). The observed $N$ photon locations are $X_i$, and we wish
to see if there is evidence that the source is offset from some
pre-determined position. As noted, there are various possibilities one
may wish to test.  One is that the source is supposed to be at one of
two definite positions, both of which we know.
Another is that the source
is either at one definite position, which we know, or it is located `somewhere else'.
Here we put a Gaussian prior on the alternative position and so know
the parameters of that prior (most interestingly, the spread).
We now deal with these cases in turn.

\ssec{Example 1}

The models $H_0$ and $H_1$ hypothesize that the $N$ data $X_i$ are
drawn from unit Gaussians of mean zero and  mean $\mu$ respectively.
The evidence ratio ${\cal E}$ is then very simple:
\[ -2 \ln {\cal E} = \sum_{i=1}^N (X_i-\mu)^2 - \sum_{i=1}^N X_i^2
= N (\mu^2 - 2 M \mu), \]
where $M$ is the mean of the $N$ samples.  Clearly $M$ will be
Gaussian, of variance $1/N$, under either hypothesis, and so the
logarithm of the evidence ratio is also Gaussian.
This means that the evidence ratio will have considerable scatter,
provided $N\mu^2 \gg 1$.

\begin{figure}
\center{\includegraphics[width=8.5cm]{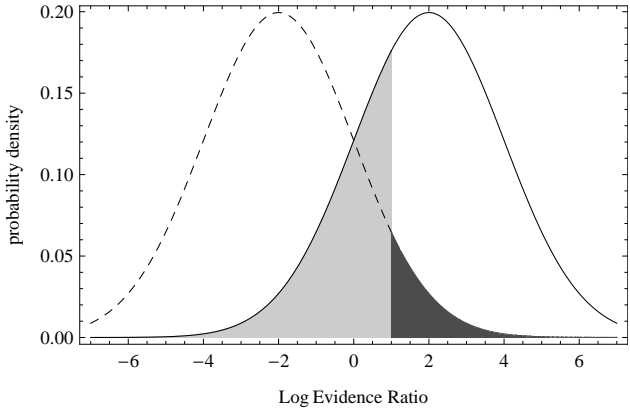} }\caption{The
  probability distributions of $\ln {\cal E}$ under $H_0$ (left) and
  $H_1$ (right), assuming $N=4$ and $\mu=1$.  For the arbitrary choice $\ln {\cal E}_c=1$,
  the dark shaded area gives the Type I error rate, and the light
  shaded area gives the Type II error rate.} \label{figure1a}
\end{figure}

\begin{figure}
\center{\includegraphics[width=8.5cm]{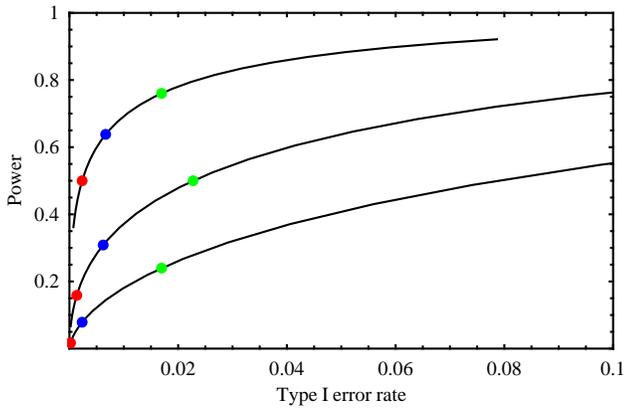}
}\caption{Parametric curves  of the power of the evidence ratio
test for example 1, as a function of the probability of Type I error, with the
parameter being the evidence ratio.  Curves are shown for $\mu=1$
and $N=2$ (lowest curve), with $N$ increasing for each curve by a
factor 2.  The dots show  evidence ratios (green, blue, red) of
$e^2$, $e^3$ and $e^4$. } \label{figure1}
\end{figure}

Our decision procedure is  to reject $H_0$ (and therefore accept
$H_1$) when  ${\cal E}$ exceeds some threshold ${\cal E}_c$.  This
occurs when $M$ exceeds
\[ M_0=\frac{\ln {\cal E}_c+N \mu^2/2}{N \mu}. \]
We make a Type I error (incorrectly rejecting $H_0$) when $M$ exceeds
this  threshold and $H_0$ is true.  The probability of this is the
probability that $M$ exceeds $M_c$ when $M$ is a Gaussian of mean
zero and variance $1/N$, which is
\[  P({\rm type\ I})=\frac{1}{2} \left( 1-\mbox{erf}\left( \frac{( 2 \ln {\cal E}_c + N \mu^2}{2\sqrt{2N} \mu} \right) \right) \]
We make a Type II error (incorrectly failing to reject $H_0$) when
$M$ is less than  $M_c$ and $H_1$ is true.  The probability of this
is
\[  P({\rm type\ II})=\frac{1}{2} \left( 1-\mbox{erf}\left( \frac{( -2 \ln {\cal E}_c + N \mu^2}{2\sqrt{2N} \mu} \right) \right) \]
Fig. \ref{figure1a} illustrates the possibilities. From these
expressions we can calculate curves of power versus Type I error
level, in which $\mu$ and $N$ are parameters. These are shown in
Fig. \ref{figure1}.

In a  classical procedure we would base our acceptance of $H_0$ on
whether $M$ differs significantly from zero.  Specifically, we
reject $H_0$ if $M > t/\sqrt{N}$ where $t$ is a parameter analogous to
${\cal E}_c$, which  we choose to determine the trade-off between power
and Type I error. We find that this test is  identical in form to
the one based on the evidence ratios (so the curves are the same in
Fig. \ref{figure1}). However, standard values of $\ln {\cal E}_c$ give very
low powers and very conservative Type I error rates, as the figure
shows. We can see why from the interesting relationship between the
two approaches:
\[ \ln {\cal E}_c =\mu \left(\sqrt{N}t-\frac{N \mu}{2} \right). \]

Suppose we design an experiment to choose between $H_0$ $(\mu=0)$ and
$H_1$ $(\mu=1)$, based on the common decisive threshold $\ln {\cal E}_c=5$.  
Once we have obtained our data and calculated the evidence
ratio ${\cal E}$, we pick $H_1$ if ${\cal E}>{\cal E}_c$ and expect
the odds on this being the correct choice to be 148 to 1.  Consider
for simplicity the case $N=1$: the critical $t$ for this example is
then 5.5 and we would need a 5.5-sigma result to reject $H_0$.
This a very conservative procedure: the Type I error rate is
$10^{-7.7}$ and the power is a paltry $10^{-5.5}$. The analysis is
warning us that setting the critical odds at the apparently desirable
148 to 1 means we will rarely exceed the evidence ratio threshold.  As
with any classical test statistic, it makes no sense to set a critical
value which will hardly ever be exceeded for the amount of data
available.  This makes it inevitable that $H_0$ will not be rejected.
We need more data, or a less stringent acceptance procedure, as Fig.
2 shows.

Suppose we carried out this experiment and obtained a single datum
that did indeed yield $t>5.5$, $\ln {\cal E}_c>5$. Formally we should
choose $H_1$ -- but common sense tells us that our datum is next to
impossible under {\em either} $H_0$ or $H_1$.  It would be more
reasonable to look for some missing hypothesis $H_2$, and certainly
prudent to check the goodness of fit of the apparently-favoured
$H_1$. We will make similar comparisons for our other examples.

%

\sssec{ROC and AUC}

The plot of power against Type I error rate is sometimes known as the
Receiver Operating Characteristic, or ROC, the name arising from its
origin in radar.  It is used a good deal in medicine (e.g.
Zweig \& Campbell 1993).  A useful quantity could be the integral
under the ROC curve, known as the AUC (Area Under the Curve). In our
application, it is not difficult to show that the AUC is the
probability that the evidence ratio, assuming $H_1$, exceeds the
evidence ratio, assuming $H_0$.  This condenses the ROC into a single
number, which for our examples is quite close to unity.  This may be a
useful compression in some cases, but it does lose the possibility of
ascribing weight to the degree by which the evidence ratios exceed
each other.

\ssec{Example 2}

In this example, the null hypothesis $H_0$ remains that the random
variables $X_i$ are drawn from a Gaussian of
mean zero and unit variance. The alternative hypothesis $H_1$ is
that the $X_i$ are drawn from a Gaussian of mean $\mu$ and unit variance.
However, in this case we do not know $\mu$ -- we assume  that the
prior on $\mu$ is Gaussian, of mean zero and known standard
deviation $\sigma$.  This formulation poses the question, `is $\mu$
zero or is it non-zero but with a restricted range of
possibilities?'  The models are nested because if $\sigma=0$ then
$H_1$ reduces to $H_0$.  Hence $H_1$ is a model that requires an extra free parameter. A model similar to this was considered by Jeffreys (1961), who examined the contrast between $p$-value and posterior probability of $H_0$.

This example shows a standard technique for creating a reasonably
comprehensive alternative hypothesis -- the use of a hierarchical
model (Gelman \& Hill 2007).  Here the introduction of one extra
parameter ($\sigma$) gives us a wide range of possible alternatives
for $\mu$, which in the previous example we had to set case-by-case.

The evidence ratio, in the sense $H_1$/$H_0$, is now
\[
{\cal E}=\frac{ \int \exp\left(-\sum_i (X_i-\mu)^2/2\right) \,
 \exp\left( -\mu^2/2 \sigma^2\right) \; d\mu }{\sqrt{2 \pi}\sigma\;
\exp\left(-\sum_i (X_i)^2/2\right). }
\]
The likelihood term depends on $\mu$ but not $\sigma$, which enters
through the prior on $\mu$.
The ratio simplifies to
\[
{\cal E}= {1\over (N \zeta)^{1/2}\sigma} \, \exp\left[\left(
{\textstyle \sum_i X_i}\right)^2/2N\zeta\right]
\label{eqn15}
\]
where
\[
\zeta \equiv 1+1/N\sigma^2,
\]
which tends to unity as $\sigma$ becomes large. Evidently, the
distribution of ${\cal E}$ under repeated trials will be determined
by the distribution of $\sum_i X_i$, which will be Gaussian under
either $H_0$ or $H_1$.

To calculate this in detail, we define:
\[
y \equiv \zeta \ln {\cal E} + {\zeta \over 2} \ln (N\zeta \sigma^2),
\label{eqn18}
\]
which becomes
\[
y=\left(\sum_{i} X_i\right)^2/\,2N. \label{yeqn}
\]
Under $H_0$, the sum  is Gaussian of mean zero and variance $N$, so
that $y$ (and hence $\ln {\cal E}$) is a $\chi^2$ variable with
one degree of freedom. Its density is
\[
dP/dy = (\pi y)^{-1/2} \exp(-y),
\]
This immediately tells us that $\ln {\cal E}$ is a $\chi^2$ variable
and so ${\cal E}$ will have considerable scatter, affecting the
power of the test.

The case of Type II error requires a little more thought.  In this
case, the $X_i$ are Gaussian with mean $\mu$, and so equation
(\ref{eqn15}) can only give us the distribution of ${\cal E}$
conditional upon $\mu$, which we do not know.  It is natural however
to marginalize over the prior on $\mu$ to obtain an unconditional
distribution for ${\cal E}$.  This is an
important conceptual step in the analysis, putting the prior spread in
a parameter, $\mu$, on the same footing as spread in the data.  The
result is that the distribution of ${\cal E}$ is broadened beyond what
would be the case if we considered repeated trials in which only
measuring error (the distribution around fixed $\mu$)
caused fluctuations in the result. We see no alternative to this
conclusion: the existence of a prior on $\mu$ means that it
must be treated as a random variable, whose value is undetermined
before we perform an experiment. The larger the uncertainty in
$\mu$, the larger the scatter in the values of ${\cal E}$ that
we can obtain.

It follows that the distribution of ${\cal E}$
depends on the distribution of $\sum X_i$ marginalized over the
prior. For $H_1$ we then find:
\[
dP/dz = (\pi z)^{-1/2} \exp(-z)
\]
with $z \equiv y/(1+N\sigma^2)$.
Returning to the Neyman-Pearson analysis, since $P({\cal E})\, d{\cal
E}=P(y)\, dy$ we can change to our convenient variable $y$ and
integrate over Gaussians to obtain
\[
 P({\rm type\ I}) = \int_{{\cal E}_c}^\infty P({\cal E}|{\rm model\ 1})\; d{\cal E}
= 1-\erf(\sqrt{y_c})
\]
and \[ P({\rm type\ II}) = \int_0^{{\cal E}_c} P({\cal E}|{\rm
model\ 2})\; d{\cal E} = \erf(\sqrt{z_c}) \] with $z_c \equiv
y_c/(1+N\sigma^2)$. The threshold $y_c$ is related to our choice of
critical ${\cal E}_c$ via equation (\ref{eqn18}).

\begin{figure}
\center{\includegraphics[width=8.5cm]{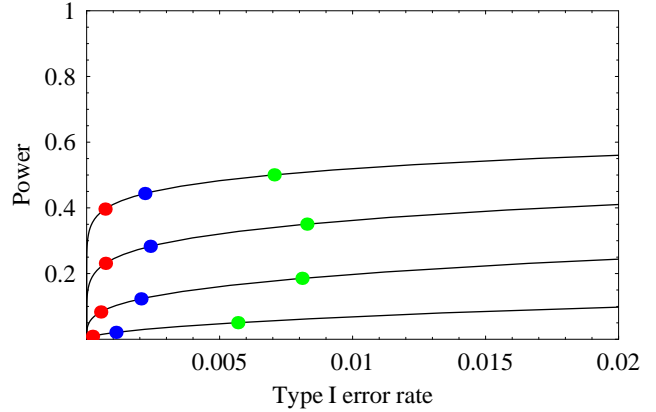}} \caption{A plot
of the power of the evidence ratio test for example 2, as a function of the
probability of Type I error.  Curves are shown for $N\sigma^2=2$
(lowest curve), increasing for each curve by a factor 2. The dots
lie at  evidence ratios of $e^2$(green),  $e^3$ (blue)  and $e^4$
(red).}\label{figure2}
\end{figure}

Fig. \ref{figure2} shows curves for the power versus
significance level as a function of  $N \sigma^2$.  This parameter
expresses the dependence of test performance on the amount of data
($N$) and the prior degree of difference between the proposed models
($\sigma$).
In this plot, it is remarkable where standard choices of critical evidence ratio lie. Take $\ln {\cal E}_c=5$ for definiteness:
at, say, $N \sigma^2=8$ we find the significance level to be $2 \times 10^{-4}$ and
the corresponding power to be $0.19$.

In words, this means that if we require the odds on $H_1$ to be 148 to
1 or stronger, then we will reject $H_0$ incorrectly only one time out
of 5000 trials, and we will pick $H_1$ when we should only one time
out of 5 trials.  This is an excessively conservative decision
procedure and parallels what we saw in the first example.

\begin{figure}
\center{\includegraphics[width=8.5cm]{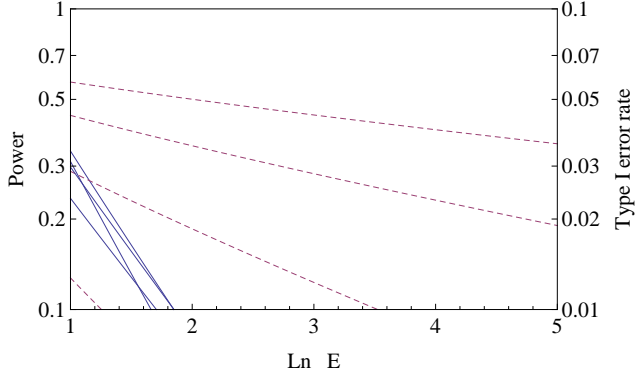}} \caption{A plot
of the power and significance levels of the evidence ratio test
for example 2, as
a function of the critical evidence ratio.  Curves are shown for
$N\sigma^2=2$ (lowest curve), increasing for each curve by a factor
2.  Significance level is  in blue, power in red. The changing
slopes of the significance lines are real.} \label{figure3}
\end{figure}

Evidently, we will need much smaller critical odds than 148 to 1 to
get reasonable performance from this test.  To re-emphasize, this is
a function of the chosen critical evidence ratio, not the form of
the test, which is intuitive.  The problem is that $\ln {\cal E}$ is
noisy for small amounts of data and cannot sustain such decisive
tests as are implied by $\ln {\cal E}_c=5$ (for example). Fig. \ref{figure3} illustrates this point.

This test might intuitively be derived without
the evidence ratio, focusing on the test statistic $(\sum_i X_i)^2$,
where the square enters to allow for the possibility that the
actual, non-zero $\mu$ can be be of either sign.  For simplicity,
consider the form $y$ we defined before, which is chi-square
distributed (see equation \ref{yeqn}). A simple test could be, reject
$H_0$ if $y>y_c$, where the critical $y_c$ corresponds to some
desired significance level or probability $p$ of Type I error. The
power of this proposed test is conditional upon $\mu$. Marginalizing
out $\mu$ with the Gaussian prior, as before, the forms of the power
and  significance level turn out to be identical to those based on
the evidence ratio test.   So the test is natural enough; the
difficulty arises in choosing a sensible value for the critical
evidence ratio.  This is exactly the same difficulty that occurs in
choosing the significance level in any classical test.

It might seem even more `natural'
for this problem to choose $\sum X_i^2$ as a test statistic. Working
through the Neyman-Pearson analysis in this case is not possible
analytically, as non-central chi-square distributions arise. However,
a numerical analysis shows that this `natural' procedure only
performs better than the evidence ratio for small $N \sigma^2<3$.
So the evidence ratio method is not trivially intuitive.

\ssec{Conclusions from the Gaussian examples}

These simple examples show that the odds that we calculate from the
evidence ratio may not be useful for making decisions if we take account of
statistical variations over an ensemble of datasets.  A decisive threshold of
$\ln{\cal E}=5$ can in many cases be exceeded only a small fraction of the
time when $H_1$ is true, even when such a value is effectively
impossible under $H_0$. In other words, the test is extremely
safe (very hard to reject the null hypothesis incorrectly),
but lacking in power (little ability to detect the alternative).
This asymmetry between type I and type II
performance seems undesirable, particularly because the problem is
set up so that there are only two possibilities.  If $H_0$ is clearly
inconsistent with the data, then $H_1$ must be correct
according to the problem as given -- even if
the Bayesian evidence ratio is only moderate. Again,
this suggests that we should be
free to challenge the statistical formulation and conclude that
neither $H_0$ nor $H_1$ are correct.  Fisher might have regarded this as the correct (less ``wooden'') approach.

\sec{A LINE FITTING EXAMPLE}

We now consider two more complex examples, where we are interested in
which of two models is a better fit to spectral line data. We will use
Monte Carlo simulation to
assess the statistical scatter between different realizations of the
data. We will consider two cases, one `nested' (whether there is an
extra component to a spectral line) and one not nested (whether a line
has a Gaussian or Lorentzian profile).

Suppose we are trying to decide if a  spectral feature is a single
Gaussian (the null hypothesis $H_0$) or two Gaussians, of equal
width, known separation, but unknown height ratio (the alternative
hypothesis $H_1$).  This is a nested model because if the height
ratio is zero, $H_1$ reduces to $H_0$. The relevant parameters are the
baseline; the height, width, and centre of the main line; and the height
ratio for the subsidiary line. The
models are:
\begin{eqnarray}
H_0: && y=\alpha_1+\alpha_2 \exp\left(-\frac{1}{2\alpha_3^2} (x-\alpha_4)^2 \right) \label{model1}\\
H_1: && y=\beta_1+\beta_2 \exp\left(-\frac{1}{2\beta_3^2}
(x-\beta_4)^2 \right)\nonumber\\
 && +\beta_5 \exp\left(-\frac{1}{2\beta_3^2}
(x-\beta_4-3 \beta_3)^2 \right) \label{model2}.
\end{eqnarray}
The extra feature is located a known three standard deviations away
from the main one. We also treat the noise levels as free parameters
to be determined from the data; this is realistic because we may not
know the noise level very well. We again assume that each model is a
priori equally likely, and that the noise is normally distributed.
The models need priors on the parameters, which we describe
later. In the Neyman-Pearson framework, our decision rule for this example will be:  accept $H_1$ if the evidence exceeds a critical value.

The Monte Carlo modelling process involves the following steps, some
repeated.

\begin{figure}
\center{\includegraphics[width=8.5cm]{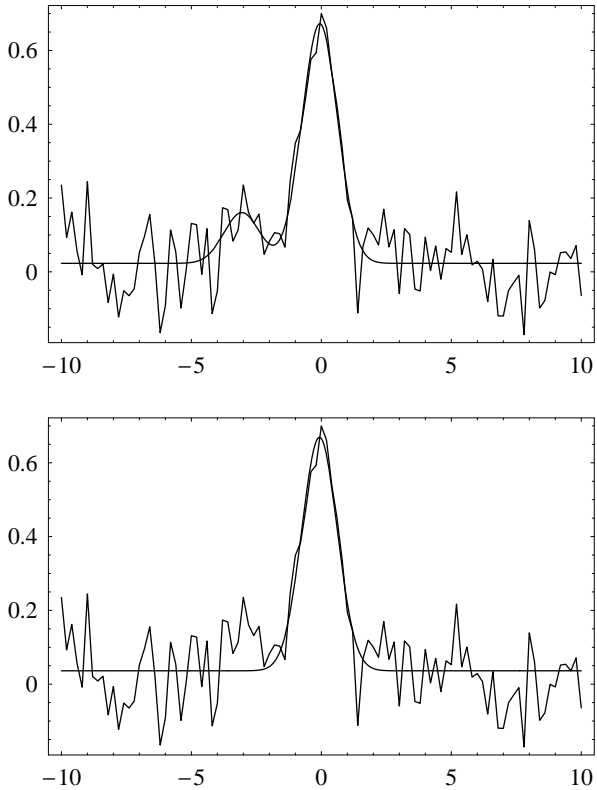}}
\caption{Both panels show fits of Gaussians, where the bimodal model $H_1$ is favoured
at odds of 20 to 1.   In the upper panel, the fit of the model under
$H_1$ is shown, in the lower, under $H_0$.} \label{figure4}
\end{figure}

\begin{enumerate}
\item To create the
noise-free spectrum under $H_0$ we take  $\alpha_1=\beta_1=0$, $\alpha_2=\beta_2=1$,
$\alpha_3=\beta_3=1$, $\alpha_4=\beta_4=0$ in equation (\ref{model1}).  The noise-free
spectra are sampled on a pixel grid of spacing $1/5$ in the above units.

\item The noise-free spectrum under $H_1$ is created with the same parameters as for $H_0$ in equation (\ref{model2}) but the key parameter $\beta_5$, the strength of the satellite line, now enters.  We will assume that the prior for $\beta_5$ is uniform
between zero and 0.1, so we are looking for a satellite
line that we expect to be at most 10\% of the main line.  The presence of this prior introduces a factor of 0.1 into the evidence ratio; we assume that the priors on the other parameters are the same for $H_0$ and $H_1$.

\item
To estimate the Type I error rate, we generate simulated data with
$H_0$ assumed true, adding Gaussian noise. We define the signal-to-noise ratio as the ratio of the
peak level (unity) to the standard deviation of the added noise.  We fit the forms for $H_0$ and $H_1$ to these
simulated  data and compute the evidence ratio in the sense ${\cal E}$
= evidence for $H_1$ / evidence for $H_0$,
using the Laplace approximation.  This gives the evidence ratio when the data are created on the assumption the $H_0$ is true.
\item
To compute the power, we need to compute the evidence ratio using data generated data for the case where
$H_0$ is false.  We will do this by assuming that $H_1$ is true.
We  generate simulated data under $H_1$ by  adding noise to the noise-free spectrum under $H_1$. Choosing the random values of
$\beta_5$, the satellite line strength, from its prior naturally
marginalizes the distribution of the evidence ratio over the range
of prior assumed strengths for the satellite line (the evidence ratio is a very strong function of this line height).
\item We fit the forms for $H_0$ and $H_1$ again and compute the evidence ratio.   Examples of
the fits under $H_0$ and $H_1$ are shown in Fig. \ref{figure4}.
\end{enumerate}

The use of the Laplace approximation is justified by examining the likelihood functions and finding them to be close to Gaussian -- a check that should always be made.

\begin{figure}
\center{\includegraphics[width=8.5cm]{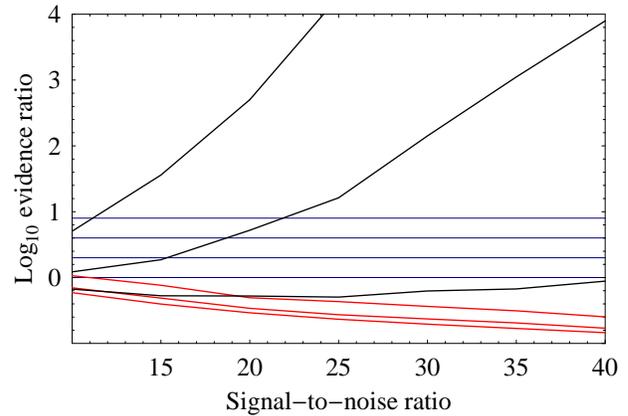}}
\caption{The evidence ratio for the first line-fitting example, in the sense evidence for
$H_1$ / evidence for $H_0$, plotted against signal-to-noise ratio.
These results include the effects of the priors that are described
in the text. The red curve assumes $H_0$ is true (no satellite line)
and the black curve assumes $H_1$ is true. The central curve is the
median and the flanking solid lines are the 25th and 75th
percentiles. 500 iterations were used at each noise level. The
horizontal lines mark odds on $H_1$ that are even, 2:1, 4:1, and
8:1.} \label{figure5}
\end{figure}

The trends of the evidence ratio with signal-to noise ratio  are
plotted in Fig. \ref{figure5}, which shows the median and the
interquartile range for the log of the evidence ratio, plotted
against the signal-to-noise ratio. We see that the evidence ratio
or odds for $H_0$, if  it is true,  do not get very big compared to
the odds for $H_1$. This is what we expect from a nested model, as
$H_1$ can always do just as well as $H_0$, with only the Ockham
penalty for extra complexity -- not severe for only one extra
parameter.  It follows from our decision rule  that the Type I error
rate is quite low. Indeed, for the standard decisive ratio of
$e^5=148$, the Type I error rates are exceedingly small -- the
decision rule is very conservative at these signal-to-noise ratios.
On the other hand, clearly if $H_1$ is true we will often find
values below the critical value and so the power is not large.
Ultimately we can trace this to the width of the prior on the
satellite line height; we are too vague about what we are looking
for to have high power. This point arises again in the next example.

\begin{figure}
\center{\includegraphics[width=8.5cm]{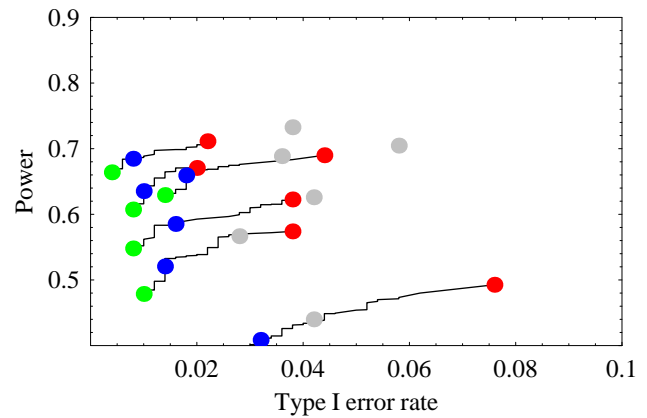}}
\caption{The power and Type I error rate for the first line-fitting
  example are shown for signal-to-noise ratios of 40 (topmost line),
  35, 30...  The critical ratio varies along each line, with points
  indicating a critical evidence ratio in favour of $H_1$ of 2 (red),
  4 (blue) and 8 (green). Grey points are for the simple case of
  picking the model with the smaller BIC. At each signal-to-noise
  ratio -- $\beta_5$ combination, 500 iterations were used.  The Type
  I error rates are therefore not reliable near $1/500$ -- the power
  should be zero for if the Type I error rate is zero, but the curves
  are too steep near the origin to resolve properly with Monte
  Carlo.} \label{figure6}
\end{figure}

The utility of the proposed decision rule is summarized in Fig.
\ref{figure6}, which shows the power and Type I error rate as a
function of decision threshold (the chosen critical evidence ratio) and
signal-to-noise ratio.  This diagram is specific to the problem at
hand, but interesting points emerge.  Evidently, the combination of
the critical evidence and the signal-to-noise ratio determines where
the decision rule places one in the diagram.   Standard decisive
thresholds like $e^5$ result in low power and a very small Type I
error rate -- less than 1/500 with our number of repetitions of the
Monte Carlo simulation. This may not be what is needed.

For comparison, we also apply a Bayesian Information Criterion (BIC; see e.g. Liddle 2007).
In our case this means we pick the model with the smallest value of
the normalized sum of squares plus the penalty term $\ln$(number of
data points) $\times$ (number of model parameters).  The number of
data points is the number of spectral channels -- evidently this
number is somewhat vague as not all channels are equally
informative.

The BIC rule, while offering no choices, sits in a useful place in the
diagram for this relatively simple problem and is no worse in power than the evidence ratio. Finally,
we note that different decision rules (for example, accepting $H_0$
if the evidence for it is bigger than the evidence for $H_1$) result
in a different diagram.

For a second example, we consider trying to decide if a line profile is Gaussian ($H_0$) or Lorentzian ($H_1$).  Here we have
\begin{eqnarray}
H_0: && y=\alpha_1+\alpha_2 \exp\left(-\frac{1}{2\alpha_3^2} (x-\alpha_4)^2 \right)
\label{model3}\\
H_1: && y=\beta_1+ {\beta_2 \over 1+ (x-\beta_4)^2 / \beta_3^2}
  \label{model4}.
\end{eqnarray}
The simulation proceeds very much as in the first case, except that
we assume the priors are the same for the two models; this is
justifiable since each of the parameters has a very similar effect
in either model. We return to the priors later. The decision rule
is, accept $H_1$ (the Lorentzian) if the evidence for it is bigger.

\begin{figure}
\center{\includegraphics[width=8.5cm]{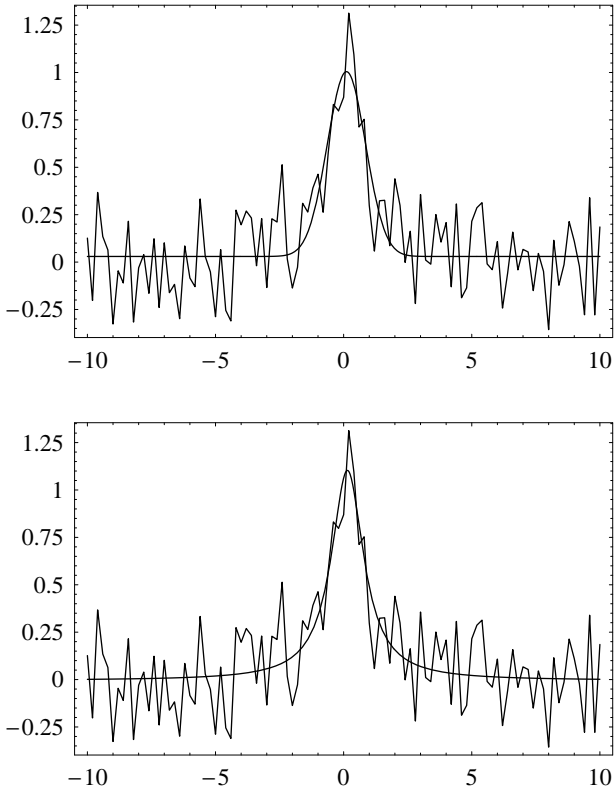}}
\caption{This shows a case where a Lorentzian (lower panel) is
favoured over a Gaussian (upper panel) at odds of 60:1. These impressive odds result from exponentiating small differences in the wings of the lines. }
\label{figure7}
\end{figure}

\begin{figure}
\center{\includegraphics[width=8.5cm]{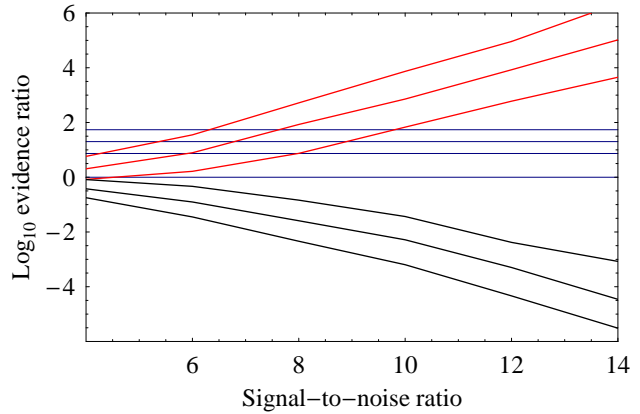}}
\caption{The evidence ratio, in the sense evidence for $H_0$ / evidence for $H_1$,
for the second line-fitting example, plotted
against signal-to-noise ratio.  The red curve assumes $H_0$ is true. The central curve is the median and the
flanking  solid lines are the 25th and 75th percentiles. 500
iterations were used at each noise level. The horizontal lines mark odds on $H_1$ that are even, $e^2$:1, $e^3$:1 and $e^4$:1.}
\label{figure8}
\end{figure}

Fig. \ref{figure7} shows examples of fits. Because the models are
not nested, and there is no Ockham factor in play, the odds (at
reasonable signal-to-noise ratios) are much stronger than in  the
previous example. Fig. \ref{figure8} shows the trends of evidence ratio with signal-to-noise.  There is much less spread in this case
because it lacks the additional variability introduced into the
previous example by the prior on the height of the satellite line.

\begin{figure}
\center{\includegraphics[width=8.5cm]{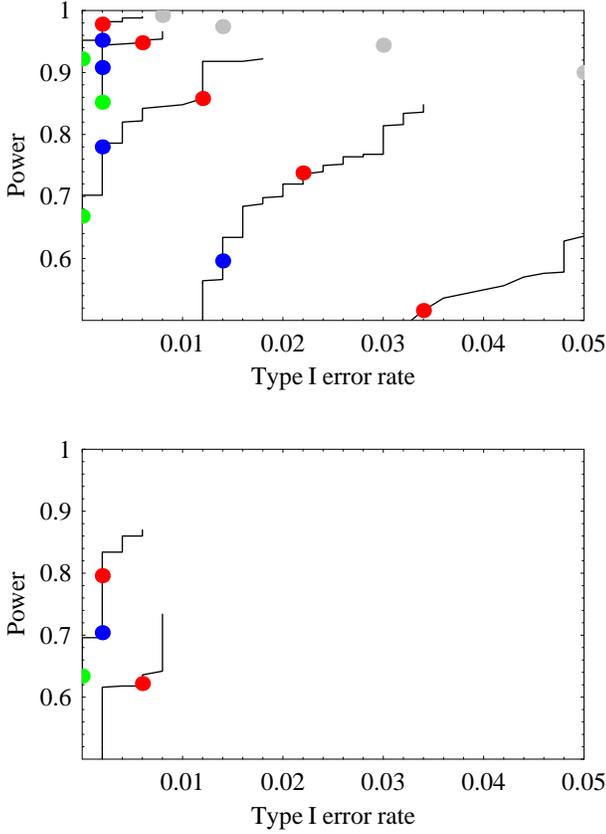}}
\caption{The power and Type I error rate for the second line-fitting example are shown for
signal-to-noise ratios  of 40 (topmost line), 35, 30... The critical
ratio varies along each line, with  points indicating critical odds
on $H_0$ of $e^2$:1 (red), $e^3$:1 (blue) and $e^4$:1 (green). Grey
points are for the simple case of picking the model  with the
smaller BIC. The upper diagram has no quadratic baseline; the lower
includes this as a possibility, as described in the text. Not all
signal-to-noise ratios yield curves that lie in the plotted ranges
of power and Type I error rate.}\label{figure9}
\end{figure}

Fig. \ref{figure9} shows the Type I error rates and powers in the
same format as before. Performance is better (lower Type I error
rate, and higher power, at the same signal-to-noise ratio),
reflecting the fact that the two models are more distinct.  Unlike the
previous example, the BIC gives better power but worse Type I error rate.

Two points emerge that are more specific to the  Bayesian context.
One is the way we have formulated the decision rule, in terms of
accepting $H_1$.  In the double-line example, it is clear than $H_1$
is the more complex model, and classically we would probably have
focused on whether or not we accepted $H_0$.  The decision rule,
`accept $H_0$ if the evidence for it is bigger than for $H_1$'
gives a different power -- Type I error rate tradeoff.  The same is
true for the second example, where this formulation (accepting the
Gaussian, in that case) gives a much higher power, and much worse
Type I error rate.

The second point relates to the role of priors.  There is no Ockham
factor at work in the second example -- but the odds that arise
seem implausibly large, just looking at the fits.  This happens
because in the simulations we fit {\it exactly\/} the right model to
the fake data.  A related point is that our model space contains
either $H_0$ or $H_1$ and nothing else. In reality it seems more
likely that we would know $H_0$, the default, null, or starting
hypothesis fairly well, but the alternative might be rather vague
Again, this reiterates the lesson of the Gaussian examples, where we
warned against adopting too restricted a set of models.

We can see the effects of this by a simple change to the Monte Carlo
modelling.  For the case $H_1$ true, we generate the fake data from
the simple Lorentzian.  We then {\it fit} a more complex model, a
Lorentzian plus a quadratic baseline $a x + b x^2$.  This
corresponds to a case where Nature is simple, but we do not know
this. Without accounting for priors, we find the odds on the
Lorentzian model drop by five orders of magnitude.  The factors
accounting for the priors on $a$ and $b$ can recover much of this.
For example, if we think in advance $a x + b x^2$ should be less
than the line height (1) over the range of the data ($\pm 10$) then
we expect the prior range in $a$ to be $\simeq 0.1$ and in $b$ to be
$\simeq 0.01$, which allows us to recover three orders of magnitude
in the evidence ratio. The second panel of Fig. \ref{figure9}
incorporates the nett loss of two orders of magnitude in the odds
for $H_1$, and we see a large effect in the power.  The Type I error
rate remains the same because we have not changed $H_0$ in any way,
reflecting its assumed role as the default, better-defined
hypothesis.  This part of the example shows how model uncertainty
may play a large role in the quality of decisions made with a
limited suite of options.

\ssec{Conclusions from the Monte Carlo examples}

The direct simulations clarify some key aspects of the evidence ratio method.  The  role of the prior is apparent; uncertainty in
our model parameters can range from sampling noise dominated (spread
in prior values $\gg$ spread due to measurement error) through to
noise-dominated.  The meta-problem of model uncertainty is also
illustrated.  A spurious restriction of the range of applicable
models is punished by equally bogus levels of certainty.  In this
regard, simple goodness-of-fit criteria have much to offer as
complements to formal procedures of model choice.

Treating the evidence ratio as a statistic shows that it has
familiar features.   Choosing a `decisive' value, in the face of
poor signal-to-noise ratio or similar models, results in
conservative decision procedures with low power to pick
alternatives.  Really this is just a reprocessing of the lack of
information in the data, but the evidence ratio encodes this fact in
an obscure way.  There is no obvious relationship between the
posterior odds, the Type I error rate, or the power:
signal-to-noise trumps all of these.

\sec{Why does the evidence ratio scatter so much?}

We have discussed at length a number of specific examples, which raise
various questions concerning the evidence ratio methodology.  The
common feature has been the large dispersion in the evidence ratio,
and the associated poor performance as a statistical tool for decision
making. We now have to ask if our examples are just pathological
cases, or if there is a general reason for the large scatter in the
evidence ratio.

Suppose we are dealing with Gaussian statistics and we are fitting
functions $y=f(x,\vec{\alpha})$ and $y=g(x,\vec{\beta})$ to data
$Y_i$, with known noise, at a set of points $x_i$. In the Laplace
approximation, the statistics of the evidence ratio are largely
dominated by the likelihood ratio
\[
{\cal L} = \frac{\exp \left(-\frac{1}{2}
\chi^2(\vec{\alpha^*}) \right) }{\exp \left(-\frac{1}{2}
\chi^2(\vec{\beta^*}) \right)},
\]
where the star superscript denotes the maximum likelihood value of
the parameters. The logarithm of this is
\[
-\frac{1}{2 \sigma^2} \left(\sum_i (f(x_i,\vec{\alpha^*})-Y_i
)^2- \sum_i (g(x_i,\vec{\beta^*}) -Y_i)^2 \right).
\]
Collecting terms, the contribution at each $i$ to the summation is
\[
\left( f(x_i,\vec{\alpha^*})^2-(g(x_i,\vec{\beta^*})^2 \right) +
\left( 2 ( f(x_i,\vec{\alpha^*})-g(x_i,\vec{\beta^*}) ) Y_i \right).
\]
If $f$ (for example) is the correct model, then $Y_i \simeq
f(x_i,\vec{\alpha^*})+Z_i,$ where $Z_i$ is a Gaussian variate by
assumption. Introducing the distance $h$ between the models by
$h(x_i,\vec{\alpha^*},\vec{\beta^*})
 =f(x_i,\vec{\alpha^*})-g(x_i,\vec{\beta^*})$
we see we have a sum of terms in which the stochastic component is
of the form
\[
\delta\ln{\cal L}= \sum_i h(x_i,\vec{\alpha^*},\vec{\beta^*})Z_i/\sigma^2,
\]
where $\sigma$ denotes the rms noise.

We therefore expect the logarithm of the evidence ratio (the log odds)
to have a Gaussian distribution, with the width of this Gaussian
determined by the detail of the distance function $h.$ There are some
special cases where this width might be small. One is where the
`incorrect' model $g$ is just the `correct' model $f$, plus some term
that is linear in the parameters. This of course would be true for
any pair of linear models where $g$ was a more complicated version of
$f$. In such a case, $h$ can be small. More generally, if $f$ and
$g$ are nested models, with a parameter that is free to be determined
in $f$ but fixed in $g$, then $h$ will also be small
if the fixed value is close to the optimum.

In general, however, apart from special cases, we should expect that
the log-odds will have a degree of variance for different realizations
of the data. The extent of the variance will depend on the details of
the models being compared but can be considerable for commonplace
problems. Since the variations in $h$ that will be consistent with
either model will naturally be $\sim \sigma$, this suggests a scatter
in $\ln{\cal E}$ of order unity, as found in our examples.

Finally we note the similarity of this treatment to the classical
likelihood ratio test. In this test, the logarithm of the ratio of
maximum likelihoods is found to be distributed like chi-square, if
the models are nested. The relevance for us is that here we have a
test statistic that is of very similar form to the evidence ratio,
and which will inevitably show considerable scatter -- although
the exact degree of scatter must be calculated in individual cases,
most simply by Monte Carlo simulation.

\section{Summary and conclusions}

We have discussed the application of the Bayesian evidence ratio to
some simple problems of model selection, which are illustrate issues encountered in astronomical applications. We have
treated the evidence ratio  as a statistic that can be computed for
a given dataset. Using analytical arguments and Monte Carlo
simulation, we have investigated the distribution of evidence ratio
values that results from an ensemble of experiments. Although the
Bayesian approach treats the data as given, our data are but one
sample of a range of possibilities. The evidence ratio calculation
indeed assumes  that we know the distribution of the data, so a
calculation  that incorporates this random element is always
possible.  All experiments could in
principle  be repeated (even the single datum of our
Universe, if we accept the Landscape picture -- e.g. Susskind 2003).

In the examples we have investigated, we find that the evidence ratio
has a large dispersion about its ensemble average, and we have given
arguments to suggest that such behaviour is likely to be general.  We
therefore suggest that one
should compute not only the value of ${\cal E}$, but also its
distribution.

Of course, at any one time, we have to work with the data
that exist, and nothing that we write here should be taken as
challenging the fundamental Bayesian paradigm: modulo the
often unjustly neglected priors on model classes, the evidence ratio does express our full knowledge of the relative odds
on different models. But when one moves beyond this statement
to a {\it decision\/} (the evidence is large enough, so I will
build a factory to produce a new drug,
or I will launch a new Great Observatory), then we need to know
more about the evidence than a single numerical value.

This point is of particular importance in calculations that
attempt to predict the potential decisiveness of future experiments,
as in e.g. Heavens et al. (2007) or Trotta (2007b). The expected
evidence ratio for a given experiment, $\langle{\cal E}\rangle$, is an
interesting quantity to know -- but one may not want to choose an
experimental strategy that maximises this quantity, if the price is an
increased scatter.

The `noisy' nature of the evidence ratio has a number of
implications. At the practical level, there is a limit to how much
effort it is worth investing in numerical algorithms for accurate
computation of the evidence ratio: arguably, there is no point in
evaluating ${\cal E}$ to better than a factor of 2 numerical precision. The
Laplace approximation may be useful here, as may judicious use of
approximate combinations of models (sometimes called `toy' models,
although this term can be an injustice).

Most seriously, the scatter in ${\cal E}$ means that it is inevitable
that a good fraction of experiments will fail to find decisive
evidence in favour of a more complex model, even when it is true (a
large `Type II' error rate, or low power). In such cases, do we accept
that the simpler model is still an acceptable description of the data?
The problem with this conclusion is that performance of the evidence ratio when viewed as a statistic seems to be asymmetric between Type I
and Type II errors. There may well exist levels of evidence that are
far from being decisively in favour of a complex model (${\cal E}\sim
1$), and are yet close to impossible on a simpler model. It such
circumstances, it hardly seems sensible to persist with the simple
model.

This reasoning is certainly reminiscent of the current controversy
over whether a scale-invariant spectrum of cosmological mass
fluctuations is ruled out (only moderate evidence in favour of tilted models
$\ln{\cal E}=2.8$, according to Trotta 2007a, even
though the observed deviation from $n=1$ is `$3.3\sigma$').  It
would be interesting to study this situation using the approach of
Monte Carlo realizations that we have advocated here, to see what
the evidence ratio is really telling us.

In such circumstances, where the evidence ratio fails to discriminate
clearly between alternative models, and yet the available data are a
poor match to the null model, we have to question whether the problem
is correctly formulated. While the null hypothesis will normally be
rather specific, the alternative can be vague, and difficult to reduce
to a single model (a point made by Efstathiou 2008 in his critique of
the evidence ratio approach).  But if the null model gives a poor fit
to the data to hand, we have good reason to believe that we need
another model, even if the the evidence ratio is unable to convince us
that this must be the supplied alternative. The strength of the
Bayesian approach is that it focuses our attention on the actual
models we are considering.  But without a goodness-of-fit test, the
result may simply demonstrate that we lack the imagination to create a
sufficiently exhaustive set of models.

\section*{Acknowledgements}

We are grateful to Mike Hobson, Andrew Liddle, Roberto Trotta and
Jasper Wall for discussions and responses to drafts of this work;
these were an important influence in helping us clarify our
presentation, insofar as this has been achieved.
Helpful comments from a referee
also led to substantial improvements in the text.

\section*{REFERENCES}

\japref Berger J.O., Sellke T., 1987.  J. Am. Stat. Assoc., 82. 112.

\japref Efstathiou G., 2008, MNRAS 388, 1314

\japref Feroz F., Hobson M.P., Bridges M., 2009, MNRAS, 398, 1601

\japref Fisher, R.A., 1956. In Chapter 4 of {\it Statistical Methods and Scientific Inference} Edinburgh: Oliver \& Boyd.

\japref Gelman A., Hill J., 2007, {\it ``Data Analysis Using Regression and Multilevel/Hierarchical Models\/}, Cambridge University Press

\japref Gregory P.C., 2005, ApJ, 631, 1198

\japref Heavens A.F., Kitching T.D, Verde L., 2007, MNRAS, 380, 1029

\japref Hobson M.P., Jaffe A.H., Liddle A.R., Mukherjee P. \& Parkinson D., 2010, {\it Bayesian methods in cosmology\/}, Cambridge University Press

\japref Jaynes E.T., 2003, {\it Probability Theory\/}, Cambridge University Press

\japref Jeffreys H., 1961, {\it Theory of Probability\/}, Oxford University Press

\japref Liddle A.R., 2004, MNRAS, 351, L49

\japref Liddle A.R., 2004, MNRAS,  377, L74

\japref Lindley D.V., 1957, Biometrika, 44, 187

\japref Niarchou A.,  Jaffe A.H,   Pogosian L., 2004. Phys. Rev. D 69, 063515

\japref Mackay D.J.C., 2003, {\it Information theory, inference, and learning algorithms\/}, Cambridge University Press

\japref Mukherjee P., Parkinson D., Liddle, A.R., 2006, ApJ, 638, L51

\japref Sellke T., Bayarri M.J., Berger J.O., 2001.  {\it The American Statistician}, 55, 62.

\japref Skilling J., 2004, in {\it Bayesian inference and maximum entropy methods in science and engineering\/}, AIP Conference Proceedings, 735, 395

\japref Susskind L., 2003, arXiv:hep-th/0302219

\japref Tegmark M., Taylor A.N., Heavens A.F., 1997, ApJ, 480, 22

\japref Trotta R., 2007a, MNRAS, 378, 72

\japref Trotta R., 2007b, MNRAS, 378, 819

\japref Trotta R., 2008,    Contemp. Phys. 49,  71

\label{lastpage}

\end{document}